\documentclass[aps,prl,reprint,groupedaddress]{revtex4-1}
\usepackage{epsfig,graphics,amssymb,amsmath,color,subeqnarray}

\usepackage[sfdefault=cmbr]{isomath} 

\DeclareMathAlphabet\mathcalbf{OMS}{cmsy}{b}{n}

\def \t{\tensorsym}
\def \lb{\left}
\def \rb{\right}

\def \d{\,\text{d}}

\def \bgamma{\boldsymbol{\gamma}}

\def \bgammad{\dot{\boldsymbol{\gamma}}}
\def \bgammadh{\hat{{\dot{\bgamma}}}}

\def \bnabla{\boldsymbol{\nabla}}

\def \bOmega{\mathbf{\Omega}}

\def \para{\parallel}

\def \bsigmah{\hat{\boldsymbol{\sigma}}}

\def \btau{\boldsymbol{\tau}}

\def \bzero{\mathbf{0}}

\def \fB{\mathcal{B}}

\def \tEh{\mathsf{\t{\hat{E}}}}

\def \tEh{\mathsf{\t{\hat{E}}}}

\def \bF{\mathbf{F}}

\def \tF{\mathsf{\t F}}
\def \tFh{\mathsf{\t{\hat{F}}}}

\def \bI{\mathbf{I}}

\def \bL{\mathbf{L}}

\def \fO{\mathcal{O}}

\def \tRh{\mathsf{\t{\hat{R}}}}

\def \tTh{\mathsf{\t{\hat{T}}}}

\def \bU{\mathbf{U}}

\def \tU{\mathsf{\t U}}
\def \tUh{\mathsf{\t{\hat{U}}}}

\def \fV{\mathcal{V}}

\def \be{\mathbf{e}}

\def \bn{\mathbf{n}}

\def \bu{\mathbf{u}}

\def \bx{\mathbf{x}}

\begin{document}
\title{Active particles in viscosity gradients}
\author{Charu Datt}\email{Present address: Physics of Fluids Group, Faculty of Science and Technology, University of Twente, 7500AE Enschede, The Netherlands}
\affiliation{Department of Mechanical Engineering and Institute of Applied Mathematics, 
University of British Columbia,
Vancouver, BC, V6T 1Z4, Canada
}

\author{Gwynn J. Elfring}\email{Electronic mail: gelfring@mech.ubc.ca}
\affiliation{Department of Mechanical Engineering and Institute of Applied Mathematics, 
University of British Columbia,
Vancouver, BC, V6T 1Z4, Canada
}
\date{\today}

\begin{abstract}{Microswimmers in nature often experience spatial gradients of viscosity. In this work we develop theoretical results for the dynamics of active particles, biological or otherwise, swimming through viscosity gradients. We model the active particles using the squirmer model, and show how viscosity gradients lead to viscotaxis for squirmers, and how the effects of viscosity gradients depend on the swimming gait of the microswimmers. We also show how such gradients in viscosity can be used to control active particles and suggest a mechanism to sort them based on their swimming style.
}
\end{abstract}

\maketitle

Cells often swim in environments, such as biofilms and mucus layers, that have spatial gradients of viscosity \citep{biofilm_viscosity,viscosity_variation}. Much like the effect of other gradients, such as light (leading to phototaxis \citep{Bennett_2015}), chemical stimuli (chemotaxis \citep{BERG_1972}), magnetic fields (magnetotaxis \citep{Waisbord_2016}), temperature (thermotaxis \citep{Bahat_2003}), or gravitational potential (gravitaxis \citep{Roberts_2010}), gradients of viscosity can lead to viscotaxis in microswimmers. Bacteria like \textit{Leptospira} and \textit{Spiroplasma}  are known to move up the viscosity gradients (positive viscotaxis) \citep{Takabe_2017, Daniel_1980}, whereas  \textit{Escherichia coli} demonstrates negative viscotaxis \citep{Sherman_1982}. It is suggested that viscotaxis plays an adaptive role in microorganisms; it prevents them from being stuck in regions where they are poor swimmers \citep{Takabe_2017, Daniel_1980, Sherman_1982}. This migration across regions of different viscosity affects organisms' population distribution, and possibly their virulence \citep{viscotaxis_petrino}.  The aggregation of microswimmers in specific regions of viscosity may also be used for sorting of cells \citep{yoshioka2006hydrogel}.

Recent works have investigated the effect of viscosity gradients on the motion of both passive and active (swimming) particles. \citet{laumann2019focusing}  have shown that viscosity gradients can be used to sort soft passive particles in microflows using cross-streamline migration. \citet{PRFStone} considered the dynamics of a hot particle in a fluid where variations in viscosity are due to the temperature difference between the particle and its surroundings. In terms of active particles, the works of \citet{Lowen_2018} and \citet{eastham2019axisymmetric} have investigated the motion of model swimmers as they move through gradients of viscosity in an otherwise Newtonian fluid.  \citet{Lowen_2018} studied the physical mechanism of viscotaxis, using assemblies of one, two and three spheres with effective propulsion forces as model swimmers, and showed how viscotaxis can emerge from a  mismatch of viscous forces on different parts of the swimmer, thereby demonstrating the possibility of both positive and negative viscotaxis i.e. motion towards or away from regions of higher viscosity, respectively. In particular, they proved that, for these model swimmers, uniaxial swimmers do not display viscotaxis. Guasto \textit{et al.} also recently implemented a systematic experimental investigation of viscotaxis with microalgae \citep{Guasto_thesis}. \citet{eastham2019axisymmetric} studied an adaptation of a common model microswimmer\textemdash  the squirmer  \citep{lighthill1951, Blake1971, Pedley_review}\textemdash as it moves through fluids in which viscosity is a function of the concentration of nutrients. They find that the coupling between viscosity and nutrient concentration can lead to qualitative differences in the swimming motion compared to fluids with constant viscosity. 

In this work, we study viscotaxis using the classical spherical squirmer model \citep{lighthill1951, Blake1971, Pedley_review}. Squirmers can be used to model different types of swimmers, from biological micro-organisms to diffusiophoretic Janus particles, within the same theoretical framework, and have been used in understanding swimming at small scales in both Newtonian (e.g., see \citep{Pedley_review} and references within)  and non-Newtonian fluids (e.g., \citep{jfmCharu, datt17, High_Deborah, Lailai, decorato, Li2014}). The squirmer model allows us to study the response of different classes of microwswimmers, namely, pushers, pullers and neutral swimmers, to spatial gradients in viscosity. We neglect any variations in viscosity gradients due to the motion of the fluid, i.e.\ we assume the P\'eclet number associated with the diffusion of the viscosity field is small \citep{eastham2019axisymmetric, PRF_shoele}. Notably we find that, in contrast to the results in \citet{Lowen_2018}, these uniaxial swimmers all display (negative) viscotaxis due to the effect of viscosity gradients on the thrust generated by the swimming gait (an effect not included in that work). We also find that the three types of swimmers behave differently in viscosity gradients, and discuss how their different swimming dynamics can be used to sort them based on their swimming style. In the following, we first present the theoretical formulation of the problem, followed by results and discussion. 

In the squirmer model a microswimmer is represented as a sphere with a prescribed surface velocity that approximates the detailed propulsion mechanism of the swimmer \citep{lighthill1951}.  We consider axisymmetric squirmers with steady tangential surface velocities $\bu^S = u^S \be_{\phi}$, where 
$
u^S = \Sigma_{l=1}^{\infty} B_l V_l \left( \phi \right),
$
and $V_l \left( \phi\right) = - 2 P_l^1 \left( \cos{\phi}\right) / \left(l \left(l+1\right) \right)$ with $P_l^1$ being the associated Legendre function of the first kind and $\phi$ the polar angle measured from the axis of symmetry of the swimmer \citep{lighthill1951, Blake1971}. The orientation of the swimmer, along the axis of symmetry is denoted by the unit vector $\be$. The coefficients $B_l$ are called the squirming modes. In fluids with constant viscosity,  the propulsion velocity of the squirmer is due to just the first mode $B_1$, whereas $B_2$  gives the strongest contribution to the flow far from the swimmer \citep{Blake1971, ishikawa, Pak2014}. Without loss of generality we take $B_1\ge 0$. For swimmers that generate thrust from the front, the puller type (like \textit{Chlamydomonas}), the ratio $\alpha = B_2/ B_1$ is greater than zero, and for those that generate thrust from the rear (like \textit{Escherichia coli}), $\alpha < 0$. When $\alpha = 0$ we have neutral squirmers. We note that a swimming gait, and thus the surface velocities of a squirmer, may be affected by changes in viscosity, but here we assume it remains fixed. Similarly to \citet{Lowen_2018}, we neglect thermal fluctuations and consider here only the deterministic motion of the squirmers.

We consider Newtonian fluids with spatial gradients of viscosity that may arise, for example, due to temperature or due to gradients of concentration of solute. We consider only small variations in viscosity to allow analytical progress \citep{PRFStone, PRF_shoele}, representing the viscosity field as 
$
\eta\left(\bx \right) = \eta_0 + \varepsilon \eta_1 ( \bx),
$
where $\varepsilon\ll1$ is a small dimensionless parameter representing the characteristic magnitude of the change in viscosity $\Delta\eta/\eta_0$. We primarily consider linear viscosity fields with slope $\eta_0/L$ (in the $\be_x$ direction for example) so that
\begin{align}
\bnabla \eta = \frac{\eta_0}{L}\be_x. \label{linearfield}
\end{align}
Provided the length $L$ is much larger than the size of the swimmer of radius $a$ such that $\varepsilon = a/L \ll 1$, the viscosity varies only weakly in the vicinity of the particle. In this case $\eta_1 =\eta_0 (x-x_0)/a$, and $\eta_0$ occurs at an arbitrary point $x_0$ in the fluid. 
 
We neglect any fluid and solid body inertia, and study microswimmers at zero Reynolds number. In the absence of inertia, the velocity of an active (or passive) particle in a fluid of arbitrary rheology may be written as 
\begin{align}\label{reci_main_1}
\tU=\tRh_{\tF\tU}^{-1}\cdot\lb[\tF_{ext}+\tF_S+\tF_{NN}\rb],
\end{align}
where $\tU=\left[\bU \ \ \bOmega\right]^\top$ is a six-dimensional vector comprising of rigid-body translational and rotational velocities whereas $\tF = \left[\bF \ \ \bL\right]^\top$ represents force and torque. The expression  \eqref{reci_main_1} can be easily derived using the reciprocal theorem of low Reynolds number flows and its formulation as described in \citep{elfring17}. For freely swimming density matched bodies the external force $\tF_{ext} = \bzero$.

The propulsive force (thrust or swim force \cite{yan15}),
\begin{align}
\tF_S = \int_{\partial\fB}\bu^S\cdot(\bn\cdot\tTh_\tU)\d S,
\end{align}
is due to any surface deformation or activity, $\bu^S$, of the particle in a Newtonian fluid of \textit{uniform} viscosity $\eta_0$. Here $\partial\fB$ represents the surface of the particle. 

The contribution due to a deviation from a Newtonian fluid with uniform viscosity, is given by
\begin{align}\label{forcenn_contri}
\tF_{NN} = -\int_{\fV}\btau_{NN}:\tEh_{\tU}\d V.
\end{align}
This term represents the extra force/torque on the particle due to the extra deviatoric stress $\btau_{NN}$ in the fluid volume $\fV$ in which the particle is immersed. The total deviatoric stress in the fluid is $\btau = \eta_0 \bgammad + \btau_{NN}$ and so in our case $\btau_{NN}=\varepsilon \eta_1\bgammad$. 

Finally, the hat quantities are linear operators from the resistance/mobility problem of a body of the same shape in a Newtonian fluid of uniform viscosity $\eta_0$, 
$
\bgammadh/ 2 = \tEh_{\tU}\cdot\tUh$,
$\bsigmah = \tTh_{\tU}\cdot\tUh$,
$\tFh = -\tRh_{\tF\tU}\cdot\tUh.
$
In this case, the rigid-body motion of a single sphere in an unbounded and otherwise quiescent Newtonian fluid, and therefore $\tEh_{\tU}, \tTh_{\tU}$ and $\tRh_{\tF\tU}$ are well known. 

The extra stress $\btau_{NN}$ requires resolution of the flow field generated by the swimmer in a fluid with a non-uniform viscosity. This task is made simpler by considering small deviations from a uniform viscosity $\varepsilon\ll 1$ and thus constructing $\btau_{NN}$ asymptotically. To study the effect of small viscosity variations, we expand flow quantities in a regular perturbation expansion of $\varepsilon$, e.g., 
$
\left\{ \bu, p, \btau \right\} = \left\{ \bu_0, p_0, \btau_0\right\}  + \varepsilon \left\{ \bu_1, p_1,\btau_1\right\}  +  \varepsilon^2 \left\{ \bu_2, p_2, \btau_2\right\} + \ldots ,
$
where $\left\{ \bu_0, p_0, \btau_0\right\}  $ are the velocity, pressure and deviatoric stress solution to Stokes equations for Newtonian fluids with uniform viscosity $\eta_0$, since at the leading order, 
$
\btau_0 = \eta_0 \dot{\bgamma}_0.
$
 At $\mathcal{O} \left( \varepsilon \right)$,
$
\btau_1 = \eta_0 \dot{\bgamma}_1 +   \eta_1 \left( \bx \right) \dot{\bgamma}_0,
$
so that, $\btau_{NN,1} =  \eta_1 \left( \bx \right) \dot{\bgamma}_0$. The extra force and torque on the particle from \eqref{forcenn_contri} then take the form
\begin{align}
\bF_{NN} &= -\varepsilon\int_{\fV} \eta_1 \left( \bx \right) \bgammad_0:\tEh_{\bU}\d V+\fO(\varepsilon^2), \label{non-newtonian_deviation1}\\
\bL_{NN} &= -\varepsilon\int_{\fV} \eta_1 \left( \bx \right) \bgammad_0:\tEh_{\bOmega}\d V+\fO(\varepsilon^2). \label{non-newtonian_deviation2}
\end{align}
We consider corrections up to $\mathcal{O}{\left( \varepsilon \right)}$ in this work. 

Note that the above expansion is valid provided that $\eta_1 /\eta_0 \sim \mathcal{O} \left( 1\right)$, and this is in principal not true for $r \sim \mathcal{O}(1/\varepsilon)$. However, for a squirmer the far field contribution, when $r \sim \mathcal{O} \left(1 /\varepsilon \right)$, is $\mathcal{O}(\varepsilon^2)$ for the extra force and $\mathcal{O}(\varepsilon^3)$ for the extra torque and can therefore be neglected as we consider corrections to Newtonian-uniform-viscosity motion up to only $\mathcal{O} \left( \varepsilon\right)$. The velocity field due to motion of a passive sphere decays more slowly than that of a squirmer and so, in principle, the far-field would contribute. However, in linear viscosity fields, the far-field contribution to the integrals at $\mathcal{O} \left( 1\right)$ is identically zero by symmetry.

\textit{Passive particles.} In order to build intuition about the influence of viscosity gradients, we first study the motion of a passive sphere in viscosity gradients. The hydrodynamic force $\bF$ and torque $\bL$ on a rigid sphere of radius $a$ moving with translational velocity $\bU$ and angular velocity $\bOmega$ in a linear viscosity field \eqref{linearfield}, is found to be, up to $\mathcal{O} \left(\varepsilon \right)$,
\begin{align}
\bF &= - 6 \pi a \eta_0 \bU +2\pi   a^3  \bnabla\eta\times\bOmega, \\
\bL &=  -8 \pi \eta_0 a^3 \bOmega  - 2 \pi  a^3 \bnabla\eta\times\bU.
\label{angular_velocity_passive}
\end{align} 
The gradient in viscosity couples the force with the sphere's angular velocity, and the torque with the translational velocity; these couplings are absent for a sphere in a Newtonian fluid with uniform viscosity. The coupling is symmetric, which may be proved by the reciprocal theorem as shown by \citet{PRFStone}. 

\citet{PRFStone} studied the case of viscosity gradients induced by a hot particle in a viscous fluid. In this instance the viscosity field may be taken to be 
\begin{align}
\eta = \eta_0\left(1 -\varepsilon \frac{a}{r}\right),\label{stonefield}
\end{align}
where $r$ is the distance from the particle of radius $a$. As a consistency check we reproduce their results using equation \eqref{non-newtonian_deviation1}.

\textit{Active particles.} We now consider the motion of active particles in viscosity gradients. We first look at the case where viscosity variations are due to the squirmer itself, following \cite{PRFStone} we take the viscosity field as shown in \eqref{stonefield} and find that the translational velocity of the squirmer is
 \begin{align}
 \bU = \frac{2 B_1}{3} \left( 1 -  \frac{\varepsilon }{12}  \right) \be, 
 \end{align}
where $\be$ is the orientation of the squirmer. This indicates that a squirmer swimming in a region of radially increasing viscosity around it swims more slowly than in a Newtonian fluid with uniform viscosity. Because $\btau_{NN}$ is linear, this result can be understood by decomposing the swimming problem into a thrust problem and a drag problem \citep{jfmCharu}. In the thrust problem, the squirmer is held fixed and the propulsive force due to its surface velocity is calculated, while in the drag problem, the drag on a passive sphere translating with the velocity of the squirmer is calculated; $\btau_{NN}$ is formed by the uniform-viscosity thrust problem, and the uniform-viscosity drag problem, respectively. Such a decomposition shows that the viscosity gradient leads to a reduction in both the thrust (by $2\pi \varepsilon \eta_0 a B_1 \textbf{e}$) and the drag (by $5 \pi/ 3 \varepsilon \eta_0 a B_1 \textbf{e}$) but because the reduction in the thrust is greater we obtain slower swimming, much like one observes in a shear-thinning fluid \citep{jfmCharu}. The angular velocity of the squirmer is zero in this case by symmetry. 

We now consider squirmers in linear viscosity fields such as \eqref{linearfield}. In this case, exact expressions for the translational and angular velocities of general squirmers can be obtained. These are, up to $\mathcal{O} \left(\varepsilon\right)$,
\begin{align}
\bU &= \bU_N-  \frac{a B_2}{5} \left(\bI -3\be \be\right) \cdot \bnabla (\eta/\eta_0), \label{equation_motion_dim_a}\\
\bOmega &=  -\frac{1}{2}\bU_N\times\bnabla (\eta/\eta_0),
\label{equation_motion_dim_b}
\end{align}
where $\bU_N = \left({2B_1}/{3}\right)\be $ is the velocity in a (Newtonian) fluid with uniform viscosity ($\bOmega_N = \bzero$). Of all squirming modes $B_l$, only the first two contribute to the translational and rotational velocities in the linear gradients considered here. Note that the gradient in viscosity affects both the translational and angular velocities of the squirmer; even an axisymmetric squirmer can now rotate due to its own motion. The rotation of the squirmer is driven by its translational velocity and is in the opposite sense to that of a passive sphere dragged along the same direction (see \eqref{angular_velocity_passive}). Equation \eqref{equation_motion_dim_a} also suggests that even with $\bU_N = \bzero$, the squirmer may swim. That symmetric squirming modes may contribute to the swimming speed was also noted by \citet{eastham2019axisymmetric} for squirmers moving in fluids in which the viscosity is a function of the concentration of surrounding nutrients and also previously shown for squirmers in shear-thinning fluids \citep{jfmCharu}

From equations \eqref{equation_motion_dim_a} and \eqref{equation_motion_dim_b}, we note that if the swimmer moves along the gradient, $\be \para \bnabla \eta$, it does not rotate ($\bOmega= \bzero$), and therefore maintains its orientation. Depending on its type of propulsion, and consequently, on the sign of $\alpha=B_2/B_1$, it can swim faster, slower or at the same speed as in a Newtonian fluid with uniform viscosity. When the particle swims down a viscosity gradient, pusher swimmers ($\alpha < 0$) swim faster, because they generate thrust (or push) at the rear which is in a fluid more viscous than that in the front. Pullers ($\alpha> 0$), on the contrary, swim more slowly as they pull on the less viscous fluid at their front. Neutral swimmers ($\alpha = 0$), which have drag and thrust centres that coincide, swim with the Newtonian speed. The dynamics for pullers and pushers reverses when swimming up a viscosity gradient. The viscosity gradient can also cause rotation of the swimmer as well as a drift velocity  when the swimming is not parallel to the gradient. The drift along the viscosity gradient depends only on the second squirming mode $B_2$, which governs whether fluid is drawn into ($B_2 < 0$) or expelled from ($B_2 > 0$) around the equator of the swimmer and causes drift up or down the gradient respectively; this motion may arise even when $\bU_N=\bzero$.

These results can all be understood by examining the effect of the viscosity gradient on thrust and drag separately. One finds that for a squirmer, the change in thrust dominates the change in drag and, consequently, the squirmer rotates in a direction opposite to that of a passive sphere (one can picture for example a small row boat in a viscosity gradient, where the differences in the viscosity between the paddles dominates the change in dynamics).

\begin{figure}
\center
\includegraphics[width = 0.44\textwidth]{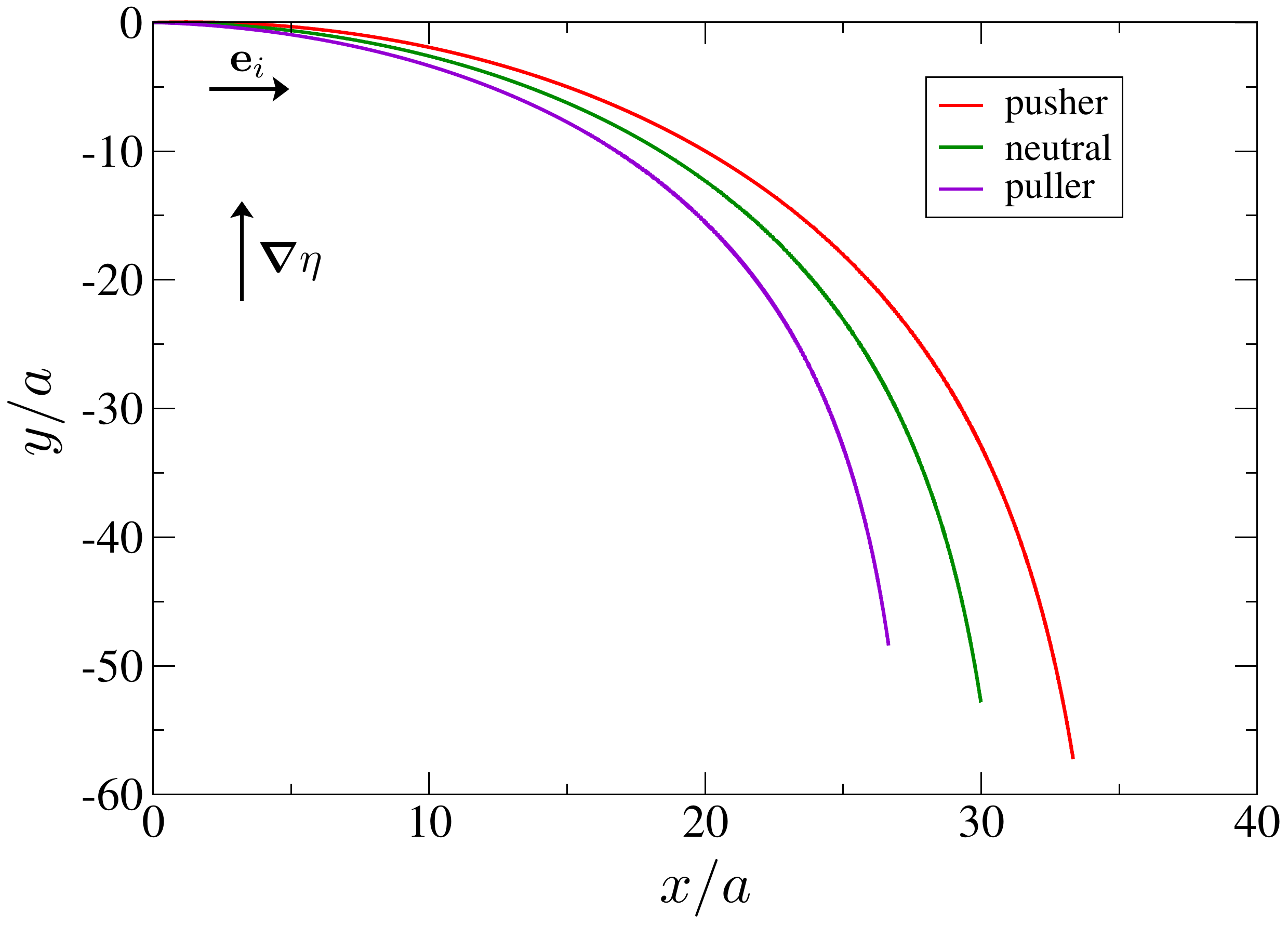}
\caption{Trajectories of squirmers with an initial orientation $\be_{i}$ orthogonal to the gradient $\bnabla\eta$, from $t  = 0$ to $t = 100a/B_1$. Eventually all squirmers swim down the viscosity gradient.}
\label{figure_linear}
\end{figure}

In general, when not perfectly aligned with the viscosity gradient, squirmers rotate toward the direction of lower viscosity and hence display viscophobicity (negative viscotaxis), irrespective of the type of propulsion. We demonstrate this in figure \ref{figure_linear}, where we consider the motion of squirmers in a linear viscosity gradient. We take $\alpha = 0$ for neutral squirmers, $\alpha = \pm 2$ for pullers and pushers, and set $\varepsilon = 0.1$.  As we can see in figure \ref{figure_linear}, the swimmers are all ultimately reoriented down the viscosity gradient with a characteristic (dimensionless) length scale of reorientation that scales as $\sim O(1/\varepsilon)$; however, the trajectories do depend on the type of swimmer. Pushers swim furthest, both laterally across the gradient and vertically along the gradient, pullers travel the least. These differences in trajectories could then be used to sort the swimmers, for example, based on distance perpendicular to the gradient, from the point of release. Non-Newtonian fluids have been demonstrated to sort swimmers based on swimming speed \cite{mathijssen16}, but here differences in swimming speed arise purely due to the effects of the viscosity gradient.

\begin{figure}[h!]
\center
\includegraphics[width = 0.45\textwidth]{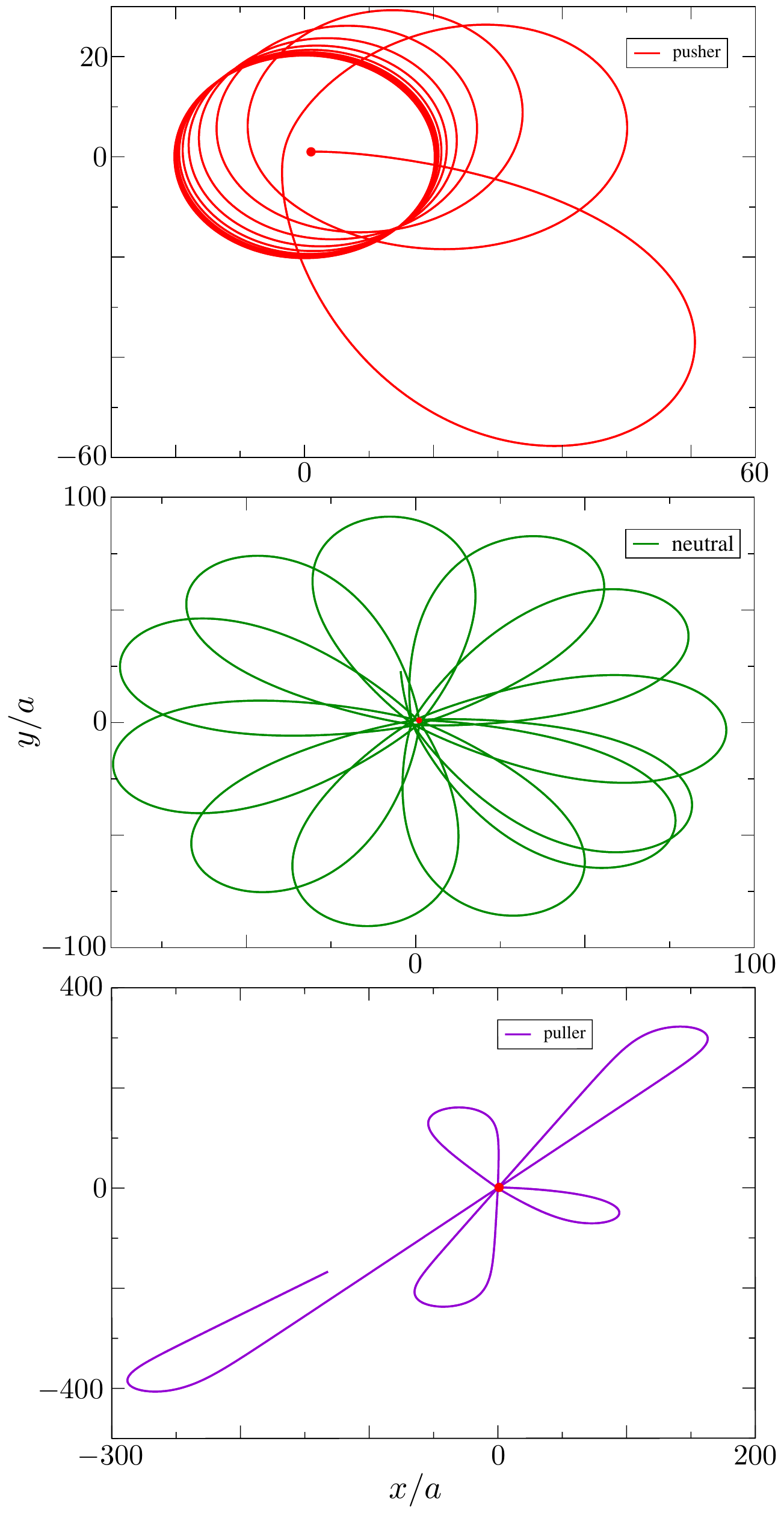}
\caption{Planar trajectories of pushers, neutral swimmers, and pullers in a viscosity field that increases radially outward from the origin, from $t=0$ to $t = 4000a/B_1$. The initial orientation of the swimmers is $\be_x$, and their initial position, $(x/a = 1,y/a = 1)$, is marked by the red dot.}
\label{figure_radial}
\end{figure}

We also plot trajectories of squirmers (in this case, with only the first two squirming modes) moving in a plane with a radially varying two-dimensional viscosity field of the form $\bnabla \eta = (\eta_0/R)\be_r$ where $R$ is a characteristic length scale describing the gradient, $\be_r$ is the cylindrical radial basis vector, and the origin in the coordinate system is defined at an arbitrary point in the fluid which may be seen as a `viscosity sink'. Here $\varepsilon = a/R \ll 1$ and $\eta_1 =\eta_0 r/a$ where $r^2=x^2+y^2$. We use a locally linear approximation whereby the local dynamics are given by \eqref{equation_motion_dim_a} and \eqref{equation_motion_dim_b} to calculate the trajectories. Although the expressions have been derived only for linear gradients, they are expected to hold for radial gradients as well, due to small curvatures at the scale of the particle (except at the origin).  The trajectories, plotted in figure \ref{figure_radial} (with $\varepsilon=0.1$ and $\alpha=0$ and $\pm2$), display qualitatively different behaviour for the three types of swimmers. Pushers find a stable orbit about the `origin' of the viscosity gradient, the trajectory of neutral swimmers is also bounded at long times but in contrast, trajectories of pullers grow unbounded. Writing the orientation vector $\be=\left\{ \cos{\theta}, \sin{\theta}\right\}$, and the viscosity gradient $\bnabla (\eta/\eta_0) = \varepsilon\left\{ \cos{\omega}, \sin{\omega}\right\}/a$, in terms of angles $\theta$ and $\omega$ measured from the $x$-axis, we can recast \eqref{equation_motion_dim_a} and  \eqref{equation_motion_dim_b} in terms evolution equations for $\psi=\theta-\omega$ and $r$. We find fixed points, when $\dot{\psi} = 0$, $\dot{r}= 0$, are 
$
r_0/a =  \left( 2+ \frac{9}{5} \varepsilon \alpha \cos{\psi_0}\right)/\varepsilon$ and
$
\psi_0 = \cos^{-1}{\left[ -(5/9 \varepsilon \alpha) \left( 1 \pm \sqrt{1 + (27/25) \varepsilon^2 \alpha^2} \right)\right]}
$ for $\alpha \neq 0$, and when $\alpha = 0$, $r_0/a= 2/ \varepsilon$ and $\psi_{0} = \pi/2$.
For pushers, the fixed point is a stable spiral, for pullers, it is an unstable spiral, for neutral swimmers, it is a centre. These characteristics can be readily observed in figure \ref{figure_radial} where we see very different dynamics for the three types of squirmers. Neutral and pusher swimmers remain bounded or trapped near the viscosity `sink', therefore puller swimmers can be sorted out at farther distances.

To estimate the potential influence of thermal fluctuations, we compare the time scale due to viscotactic reorientation from \eqref{equation_motion_dim_b}, $\tau \sim a/\varepsilon U_N$, with the rotational diffusion time scale of a sphere, $\tau_R  =  8\pi \eta_0 a^3/k_B T$, and find that thermal fluctuations can be neglected provided $k_B T /\varepsilon\eta_0 a^2 U_N \ll 1$. This implies that in a water-like fluid at room temperature thermal fluctuations are only important for micron-sized (and smaller) swimmers \citep{Lowen_2018}.

In summary, we observe that axisymmetric squirmers are, in general, viscophobic. Unless perfectly aligned with the gradient, they turn towards regions of lower viscosity. Using the squirmer model, we find that the effects of viscosity gradients on the dynamics of active particles depend on the details of their propulsion type and even simple viscosity fields lead to different dynamics for different microswimmers. The change in the dynamics of the swimmers can be readily explained by separately investigating the change in thrust and drag; while viscosity gradients affect both the drag on the body and the propulsive force generated by the swimming gait, the change in the latter tends to dominate the response for spherical squirmers; the opposite might be true for swimmers of a different shape thereby rendering positive viscotaxis. Finally, we showed how viscotaxis may be used as a method to control the motion of active particles through different viscosity profiles and highlighted how changes in the dynamics of different swimmers may be used as a mechanism to sort them based on their swimming style.

The authors gratefully acknowledge funding from the Natural Sciences and Engineering Research Council of Canada (NSERC).

\bibliography{refs}

\end{document}